\documentclass[aps,pra,preprint,superscriptaddress,showpacs]{revtex4}
\usepackage{epsfig}
\usepackage{bm}

\def\ketm#1{  \left\vert  #1   \right\rangle   }

\def\mem#1#2#3{  \left\langle #1 \left\vert  #2 \right\vert #3 \right\rangle   }

\begin{document}

%
%

\title{Relativistic calculation of the
two-photon decay rate of highly-excited ionic states}

%
%

\author{Ulrich D. Jentschura}\email{ulj@mpi-hd.mpg.de}
\affiliation{Max-Planck-Institut f\"ur Kernphysik, 
Postfach 103980, D-69029 Heidelberg, Germany}
\affiliation{Institut f\"{u}r Theoretische Physik, Philosophenweg 16, 69120 Heidelberg}
\author{Andrey Surzhykov}
\affiliation{Max-Planck-Institut f\"ur Kernphysik, 
Postfach 103980, D-69029 Heidelberg, Germany}
\affiliation{Physikalisches Institut der Universit\"{a}t, Philosophenweg 12, 69120 Heidelberg}

\date{\today}

%
%

\begin{abstract}
Based on quantum electrodynamics, 
we reexamine the two-photon decay of one-electron atoms. 
Special attention is paid to the calculation of the (two-photon)
total decay rates which can be viewed as the imaginary part of the 
two-loop self-energy. We argue that our approach can easily be 
applied to the cases with a virtual state having an
intermediate energy between the initial and the final state of the
decay process leading, thus, to the resonance peaks in the two-photon
energy distribution. In order to illustrate our approach, we obtain 
fully relativistic results, resolved into electric and magnetic
multipole components, for the two-photon decay rates of the 
$3S_{1/2} \to 1S_{1/2}$ transition in neutral hydrogen as well as 
in various hydrogen-like ions.
\end{abstract}


\maketitle
%
%

\section{INTRODUCTION}

Since the seminal work of 
G\"oppert-Mayer \cite{GoM31}, two-photon
decay rates of excited states
in hydrogen-like atoms and ions have been the subject of intense
experimental \cite{LiN65, MaS72, PeD85, MoD04} and 
theoretical \cite{BrT40,DrG81,FlE84,CrT86,FlS88,SaP98,Jen07,Jen07b} 
studies. For many years, the investigations 
have dealt not only with the total decay rates but also with the energy 
and even angular distributions of the two emitted photons. By analyzing these
(two-photon) properties, unique information has been obtained about the
structural properties of one-electron systems including subtle relativistic
effects as well as about the basic concepts of quantum 
physics such as, e.g., the entanglement.

Even though large experimental and theoretical efforts have been 
undertaken in the past to understand various aspects of the two-photon
decay of hydrogen-like atoms, the analysis of this process still raises a 
number of unresolved problems. One of these problems, which currently attracts 
a lot of interest, concerns those two-photon transition from highly excited 
states to the ground state which pass through 
an intermediate state with a lower energy than the 
initial state of the two-photon process~\cite{CrT86,FlS88} and 
can alternatively decay to the ground-state via two 
(or more) sequential one-photon emissions. 
Such a transition leads to resonance peaks in the energy spectrum 
of the coherently emitted photons from the 
two-photon decays which are located at the energies corresponding to
the (real) intermediate states. One of the most pronounced examples of such a 
situation is the $3S_{1/2} \to 1S_{1/2}$ two-photon $E1E1$ transition for which 
the differential (in energy) emission rate has singularities at energies 
corresponding to the $3S_{1/2} \to 2P_{1/2} \to 1S_{1/2}$ and
$3S_{1/2} \to 2P_{3/2} \to 1S_{1/2}$ cascades. A proper treatment of these 
singularities is obviously required for computing total decay rates obtained 
after an integration over the energies of the coherently 
emitted photons in the direct two-photon decay $3S_{1/2} \to  1S_{1/2}$. 

During the last two decades, the theoretical treatment of the resonances 
in the energy distribution of the emitted photons has been discussed in a 
number of places. In general, the decay rate $\Gamma_{i}$ of an initial
state $|i\rangle$ in a hydrogenlike atom is the sum 
of a one-photon decay rate $\Gamma^{(1)}_{i}$ and a two-photon 
contribution $\Gamma^{(2)}_{i}$,
\begin{equation}
   \label{gamma_total}
   \Gamma_{i} = \Gamma^{(1)}_{i} + \Gamma^{(2)}_{i} \, .	
\end{equation}
The expression for $\Gamma^{(2)}_{i}$ as originally derived 
in~\cite{GoM31} is easily seen to involve an integral over the 
energies of the emitted photons, the sum of which has to be  
equal to the energy difference of the initial and final states,
and a summation over all possible intermediate, virtual states.
In order to avoid problems with 
non-integrable singularities, the authors of~\cite{CrT86}
have attributed $\Gamma^{(2)}_{i}$ only to the so-called non-resonant 
intermediate transitions, in contrast to a summation over the 
complete intermediate-state spectrum. The non-resonant transitions are given by 
intermediate states of energy higher than the 
energy $E_i$ of the initial state \cite{CrT86} (the ``resonant'' intermediate states, 
which are involved in the one-photon cascade processes,
are explicitly excluded from the sum over 
intermediate states). Based on this assumption, the non-resonant contribution
for the $3S_{1/2} \to 1S_{1/2}$ two-photon transition was
calculated as $\Gamma^{(2)}_{3S} = 8.2196 \, s^{-1}$ for hydrogen. Later, this
result has been also confirmed in the calculations by Florescu
and co-workers \cite{FlS88} who used a different method
for the summation over the ``non-resonant'' states. 

Although the results presented in Refs.~\cite{CrT86,FlS88}
are in mutual agreement, 
they are both based on the interpretation of the
two-photon decay rate $\Gamma^{(2)}_{i}$ as a rate 
generated only by non-resonant intermediate levels. In our manuscript,
we would like to propose an alternative way for the computation of the
(two-photon) total decay rates which leads to a 
natural removal of the infinities otherwise introduced into the 
expression for the two-photon decay rate. We apply here a fully 
relativistic, quantum electrodynamical approach to re-investigate 
the two-photon decay of highly-excited states of hydrogen-like atoms,
paying special attention to a careful handling of the 
resonances infinitesimally displaced from Feynman's photon integration 
contour (these singularities 
exactly correspond to the problematic ``resonant'' intermediate states). 
By making use of this approach, we obtain finite, 
physically sensible results for the decay rates of the two-photon 
$3S_{1/2} \to 1S_{1/2}$ transitions in neutral hydrogen as well as
in the various hydrogen-like ions. Apart of the leading, electric
dipole ($E1E1$) transition, we also discuss the contributions
from the higher multipole components to the total decay rate. 

This paper is organized as follows: after a brief survey of the 
theoretical expressions used in our analysis (Sec.~\ref{theory}),
we proceed by discussing the method of evaluation (Sec.~\ref{evaluation})
as well as the numerical results obtained for the differential as 
well as the total two-photon decay rates (Sec.~\ref{results}).
Conclusions are given in Sec.~\ref{conclusions}.

%
%

\section{THEORY}
\label{theory}

Within quantum electrodynamics,
the (negative) imaginary part of the self-energy is just the $\Gamma/2$, 
where $\Gamma$ is the decay width \cite{BaS78,BaS91,Jen04}. 
The one-photon decay $\Gamma^{(1)}$ rate is obtained 
from the imaginary part of the one-loop 
self-energy, while the two-loop self-energy gives rise to the 
two-photon decay rate $\Gamma^{(2)}$. 
Because the relativistic formulation of the two-loop self-energy
problem has been discussed before in a number of places \cite{FoY73,YeI03},
we only mention here that by following a straightforward
generalization of the standard procedure
described for the non-relativistic framework in Refs.~\cite{Jen07, Jen07b,Jen04},
we obtain the following expression for the two-photon decay rate 
($\hbar = c = \epsilon_0 = 1$),
\begin{equation}
   \label{gamma_2_QED}
   \Gamma^{(2)}_i = \frac{\alpha^2}{\pi} 
   \lim\limits_{\epsilon \to 0} {\rm Re} \,
   \int\limits_{0}^{\omega_{\rm max}}{\rm d}\omega_1 \,
   \omega_1 \, \omega_2 \, \int {\rm d}\Omega_1 {\rm d}\Omega_2 \,
   S_{if}(\omega_1, \omega_2) \, ,
\end{equation}
where $\omega_1 + \omega_2 = \omega_{\rm max} = 
E_i - E_f$ with the initial and final state energies
$E_i$ and $E_f$, respectively. $S_{if}$ is given by
\begin{eqnarray}
   \label{amplitude_general}
& & S_{if}(\omega_1, \omega_2) = \sum_{\nu}
   \biggl( \frac{\mem{{\psi_{f}}}{\mathbf{A}^*_1}{\psi_\nu}
   \mem{\psi_\nu}{\mathbf{A}^*_2}{\psi_{i}}}{E_{i} - E_{\nu} 
   - \omega_2 + i \epsilon}
   \nonumber \\
   & & \qquad + \frac{\mem{{\psi_{f}}}{\mathbf{A}^*_2}{\psi_\nu}
   \mem{\psi_\nu}{\mathbf{A}^*_1}{\psi_{i}}}{E_{i} - E_{\nu} 
   - \omega_1 + i \epsilon} \biggr) 
   \nonumber \\
   & & \times \sum_{\rho}
   \biggl( \frac{\mem{{\psi_{i}}}{\mathbf{A}_1}{\psi_\rho}
   \mem{\psi_\rho}{\mathbf{A}_2}{\psi_{f}}}{E_{i} - E_{\rho} 
   - \omega_1 + i \epsilon}
   \nonumber \\
   & & \qquad + \frac{\mem{{\psi_{i}}}{\mathbf{A}_2}{\psi_\rho}
   \mem{\psi_\rho}{\mathbf{A}_1}{\psi_{f}}}{E_{i} - E_{\rho} 
   - \omega_2 + i \epsilon} \biggr) \, ,
\end{eqnarray}
where in the second factor, the initial and the final state
are exchanged, but the infinitesimal imaginary part in the 
denominators remains $ + i \epsilon$ (i.e., does not change sign).
We here manifestly assume that 
$\psi_{i}(\bm{r}) \equiv \psi_{n_i j_i \mu_i}(\bm{r})$ and
$\psi_{f}(\bm{r}) \equiv \psi_{n_f j_f \mu_f}(\bm{r})$ are the well-known
solutions of the Dirac Hamiltonian for a single electron in the 
standard representation, describing an electron bound
to a point-like nucleus with
charge number $Z$. For photons propagating with wave vector $\bm{k}_i \;\, (i=1,2)$ and
unit polarization vector ${\bm u_{\lambda_i}}$
($\bm{k}_i \cdot {\bm u}_{\lambda_i} = 0$), moreover, the electron-photon
interaction operator $\mathbf{A}_i$ in the transition amplitude
(\ref{amplitude_general}) can be written in velocity gauge as:
\begin{equation}
   \label{interaction_operator}
   \mathbf{A}_i = A_0 \, \bm{\alpha} \cdot \bm{u}_{\lambda_i}
   {\rm e}^{i \bm{k}_i \bm{r}} \, ,
\end{equation}
where $A_0$ is a normalization factor, ${\bm \alpha}$ are the
standard Dirac matrices, and $\lambda_i = \pm 1$ denotes
the \textit{helicity}, i.e.\ the
spin projection of the photon onto the direction of propagation $\bm{k}_i$.
It is important to note that even though the electron-photon
interaction operator (\ref{interaction_operator}) depends, of course,
on the direction of the photon emission, one has to integrate
over these directions in Eq.~(\ref{gamma_2_QED}) in order to get
the total decay rate.

\section{EVALUATION}
\label{evaluation}

The summation 
over the intermediate states in the amplitude (\ref{amplitude_general}) 
runs over the \textit{complete} one-particle spectrum 
$\ketm{\psi_{\nu}} \equiv \ketm{\psi_{n_\nu j_\nu \mu_\nu}}$, including a 
summation over the discrete part of the spectrum as well as an integration 
over the positive and negative-energy 
continuum of the Dirac spectrum. One has to use the 
full Dirac--Coulomb Green function---which is not known 
in closed analytic form---in order to perform 
this calculation consistently. 
In recent years, the Green's function method~\cite{DrS91} has been widely
applied for the analysis of the total two-photon decay rates as
well as the photon-photon angular correlation functions \cite{SuK05}.
Various possibilities for the numerical implementation of the
relativistic Green's function are known, among which we would 
like to mentionas
(i)~a well-known formulation in terms of Whittaker
functions~\cite{PJM74} and (ii) a Sturmian decomposition in 
terms of Laguerre polynomials as suggested by Hylton and 
Snyderman~\cite{HyS97}.

\begin{figure}[t]
\begin{center}
\includegraphics[width=1.0\columnwidth, angle=0]{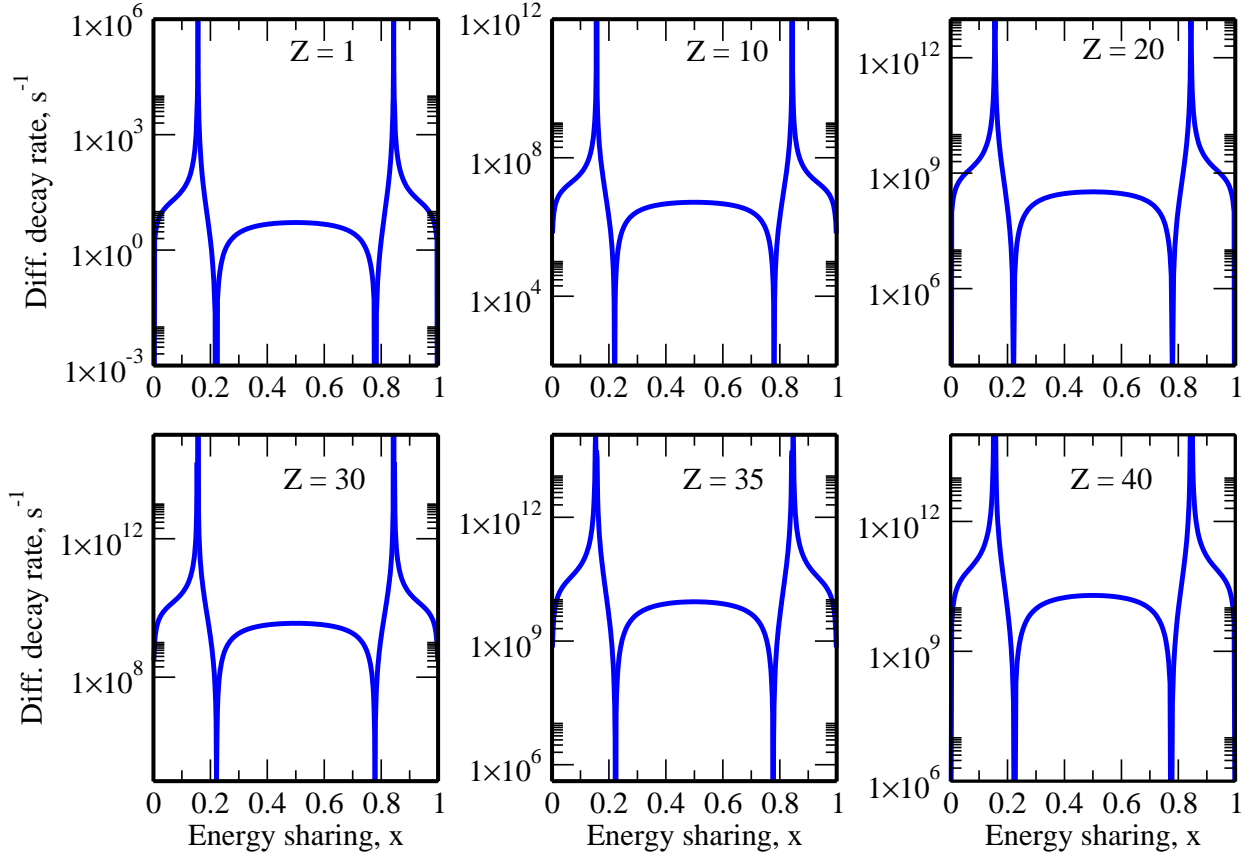}
\end{center}
\vspace*{-0.5cm} \caption{(Color online) Differential decay rate 
${\rm d}\Gamma/{\rm d}x$ for the dominant E1E1 component of the 
$3S \to 1S$ two-photon 
transitions in neutral hydrogen as well as in hydrogen-like
ions with $Z = 1, 10, 20, 30, 35$ and $40$, where $x$ is the 
energy sharing $x = \omega_1/(\omega_1+\omega_2)$.
Relativistic wave functions are used for the initial, intermediate 
and final states, but the electron-photon interaction has been restricted to 
electric dipole term ($E1 E1$ term). The resonance peaks in the decay rate
correspond to the resonant $3S_{1/2} \to 2P_{1/2} \to 1S_{1/2}$ and
$3S_{1/2} \to 2P_{3/2} \to 1S_{1/2}$ decay processes.}
\end{figure}

We use the latter representation and apply the techniques of 
Racah's algebra to all 
spherical tensors and to the standard radial-angular representation of the wave
functions, and to the Dirac--Coulomb Green function. 
For the interaction of electrons with the radiation 
field, the spherical tensor components are obtained from the known standard 
multipole expansion of the photon operator [see, e.g., Eq. (5) of 
Ref.~\cite{eic98}],
\begin{eqnarray}
   \label{photon_operator_decomposition}
   \bm{u}_{\lambda}{\rm e}^{i \bm{k} \bm{r}} =
   \sqrt{2 \pi} \sum\limits_{L M p} i^L (i \lambda)^p 
   \sqrt{2L + 1}
   \bm{A}_{LM}^{(p)} \, D^{L}_{M \lambda}(\bm{n}) \, ,
\end{eqnarray}
where $\bm{A}_{LM}^{(p)}$ denotes the electric ($p$ = 1) and magnetic ($p$ = 0)
multipole fields, respectively. 

%
%
\section{RESULTS}
\label{results}

The great advantage of the multipole decomposition 
(\ref{photon_operator_decomposition}) is that is allows us 
to study the contributions to the total (two-photon) decay rate 
from the various \textit{allowed} multipole combinations. We use
the integrand in the integral over $\omega_1$ in 
Eq.~(\ref{amplitude_general}) as a measure of the differential decay
rate (where we can set explicitly $\epsilon = 0$ for the differential 
rate). The energy distributions of the two photons emitted in the              
$3S_{1/2} \to 1S_{1/2}$ decay of neutral hydrogen and hydrogen-like 
ions are calculated as a function of the energy
sharing parameter $x = {\omega_1}/({\omega_1}+{\omega_2})$.
For an energy sharing in the range $0 < x < 1$, the contributions to 
the energy distribution from the $E1 E1$ and $E1 M2$ multipole combinations 
are displayed in Figs.~1 and 2, respectively.
As seen from these figures, the photon energy distributions
for both multipole combinations exhibit sharp resonance peaks. 
As already mentioned, this behaviour is due to the fact that the summation in 
Eq.~(\ref{amplitude_general}) includes also intermediate 
states $\ketm{\nu}$ having an energy $E_{\nu}$ with $E_i > E_{\nu} >E_f$.
However, the intermediate states contributing to the peaks 
are not only defined by the (symmetry of)
the initial $\ketm{i}$ and final $\ketm{f}$ states but also by 
the multipole components of the radiation field involved in the 
two-photon process and are different for 
$E1 E1$ as opposed to $E1 M2$, and in addition, 
marked differences exist between the low-$Z$ and the high-$Z$ region. E.g., 
the fine-structure of the 
resonance in the $E1 E1$ energy spectrum grows with $Z$
(the contributing states are $2P_{1/2}$ and $2P_{3/2}$). 
By contrast, no splitting is observed---even for very heavy ions---for the 
$E1 M2$ component of the $3S_{1/2} \to 1S_{1/2}$ decay.
Only one intermediate state, namely $2P_{3/2}$, is allowed for $E1 M2$.

\begin{figure}[t]
\begin{center}
\includegraphics[width=1.0\columnwidth, angle=0]{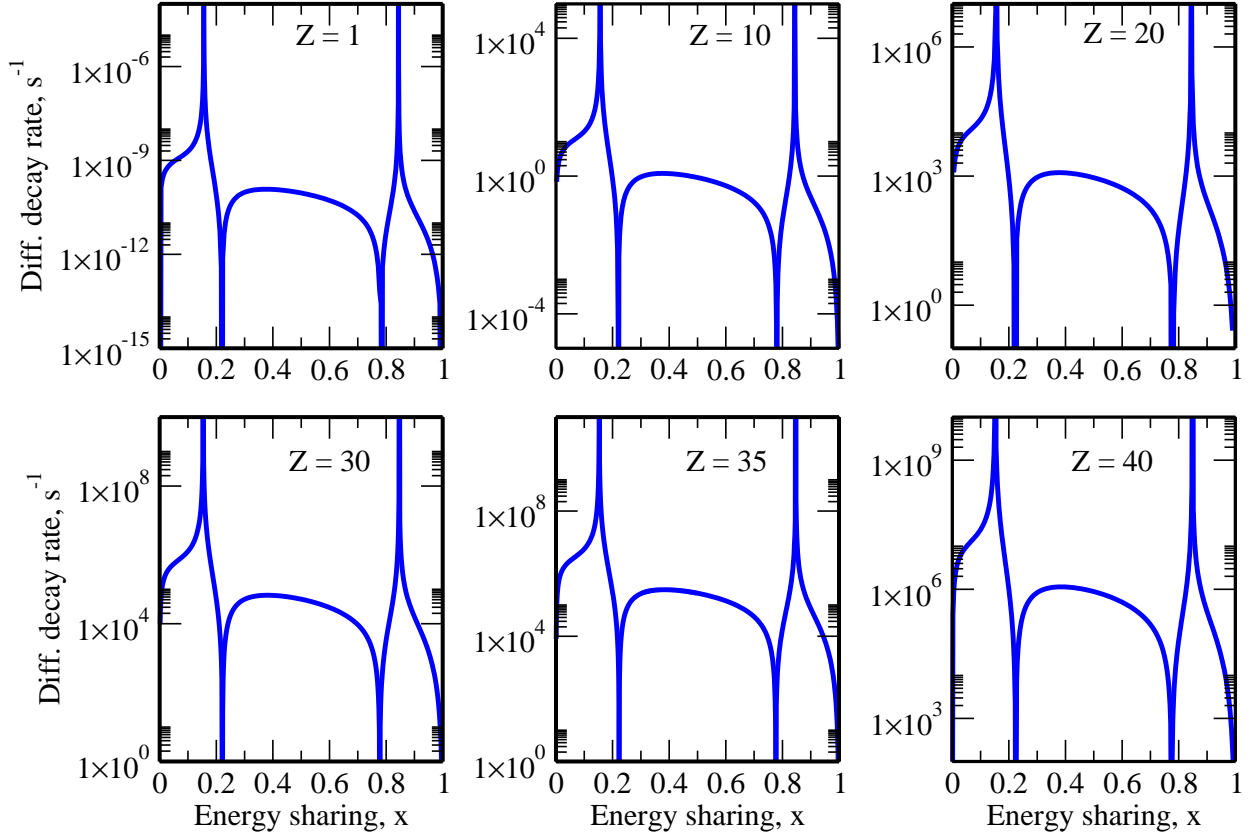}
\end{center}
\vspace*{-0.5cm} \caption{(Color online) 
Differential decay rate ${\rm d}\Gamma/{\rm d}x$
for the E1M2 component of the $3S \to 1S$
two-photon transitions in neutral hydrogen as well as in hydrogen-like
ions with $Z = 1, 10, 20, 30, 35$ and $40$ 
($x$ is the fraction of energy carried by the E1 photon). 
The $E1 M2$ term is treated relativistically. The resonances in the decay rate
exclusively correspond to the $3S_{1/2} \to 2P_{3/2} \to 1S_{1/2}$ cascade, the 
$2P_{1/2}$ state does not contribute.}
\end{figure}

Our treatment of the intermediate state resonance peaks 
in the integration over the photon energy 
is dictated by an accurate analysis of 
Eqs.~(\ref{gamma_2_QED})--(\ref{amplitude_general})
as obtained form the relativistic two-loop self energy.
The general structure of the contribution to $
S_{if}(\omega_1, \omega_2)$ generated by 
virtual states with intermediate energies $E_i > E_{\nu} > E_f$
with resonance energy $\omega_r = E_i - E_{\nu}$ or 
$\omega_r = E_{\nu} - E_f$ is as follows:
\begin{eqnarray}
   \label{resonant_terms}
   S_{if}(\omega_1, \omega_2) &\sim&
   \frac{R_1}{\omega_r - \omega + i \epsilon}
   + \frac{R_2}{\left(\omega_r - \omega + i \epsilon \right)^2} \, .
\end{eqnarray}
The integration of the first term can be carried out
using the Dirac prescription (see, for example, 
Ref.~\cite{Jen07}). The second term of Eq.~(\ref{resonant_terms}) can 
be treated using the formula
\begin{equation}
   \label{quadratic_term_integration}
   \lim\limits_{\epsilon \to 0} {\rm Re} \,
   \int\limits_{0}^{1} {\rm d}\omega
   \left(\frac{1}{\omega_r - \omega + i \epsilon} \right)^2 =
   \frac{1}{\omega_r (\omega_r - 1)} \, ,
\end{equation}
where we used an appropriate scaling of the photon energy integration
variable in order to map the integration region
to the interval $(0,1)$. 
It is important to note that Eq.~(\ref{quadratic_term_integration})
holds strictly for $0 < \omega_r < 1$, but the limit is
not approached uniformly~\cite{Jen07,Jen07b}; i.e.,~it 
would be forbidden to exchange
the sequence of the limit $\epsilon \to 0$ with the 
integration over $\omega$. As usual in quantum electrodynamic processes, 
all regulators have to be kept
up until the very end of the calculation.

With these preparations, it is easy now to integrate
over the photon energies (see Table 1 
for the $3S_{1/2} \to 1S_{1/2}$ process).
As seen from the table, the cross sections for the
$E1 E1$, $E1 M2$ and $M1 M1$ components
components scale with the nuclear charge as $Z^6$, $Z^{10}$ 
and $Z^{10}$, respectively. As expected, this scaling  behaviour
is similar to the $Z$-scaling of the multipole components
in the $2S_{1/2} \to 1S_{1/2}$ transition.

\begin{table}[t]
\vspace*{1cm}
\begin{tabular}{c@{\hspace*{0.5cm}}c@{\hspace*{0.5cm}}c@{\hspace*{0.5cm}}c} 
\hline
\hline
$Z$ & $E1E1$     & $E1M2$                & $M1M1$ \\
    & (Z$^{-6}$) & (Z$^{-10}$ 10$^{10}$) & (Z$^{-10}$ 10$^{12}$) \\
\hline
1   & 2.08  & 1.19  & 6.13 \\
5   & 2.03  & 1.18  & 6.13 \\
10  & 1.98  & 1.16  & 6.14 \\
15  & 1.94  & 1.14  & 6.16 \\
20  & 1.90  & 1.12  & 6.20 \\
25  & 1.84  & 1.08  & 6.24 \\
30  & 1.79  & 1.03  & 6.30 \\
35  & 1.67  & 0.96  & 6.39 \\
40  & 1.60  & 0.86  & 6.50 \\ 
\hline 
\hline 
\end{tabular}
\caption{\label{table1} Contributions from different combinations of multipoles
to the integrated decay rate $\Gamma^{(2)}$, in units of radians per second. 
Relativistic calculations have 
been performed for different hydrogen-like ions.}
\end{table}

Furthermore, as seen from Table 1 and as implied by the 
non-uniform convergence of the integrals, 
the intermediate states with the energies $E_\nu$ lying between 
the energies of the initial and the final states give a 
\textit{finite} contribution to the two-photon decay rate. 
For the electric dipole ($E1 E1$) transition in a neutral 
hydrogen atom, e.g., a proper treatment of the 
intermediate $2P_{1/2}$ and $2P_{3/2}$ states leads to the 
decay rate of $\Gamma^{(2)}_{3S} = 2.08 \, s^{-1}$ which is in
agreement with the result of nonrelativistic 
calculations reported in Ref.~\cite{Jen07}. However, when comparing 
our prediction with the theoretical data by Cresser and co-workers \cite{CrT86} 
a large discrepancy by about a factor of $4$ is observed.
The occurrence of the discrepancy is natural because the problematic virtual states
with intermediate energies are treated differently in \cite{CrT86}.

%
%

\section{CONCLUSIONS}
\label{conclusions}

The two-photon decay of hydrogen-like ions
has been re-investigated within the framework of relativistic
quantum electrodynamics. Starting from first principles of
this theory, we treat the total (two-photon) decay rate as the
imaginary part of the relativistic two-loop self-energy. 
The great advantage of this approach, which has its
roots in field theory, is that it provides a
simple and efficient route to handle the potentially problematic
cases of those two-photon transitions from an excited 
into the ground state which pass intermediate states that 
can otherwise also be reached in one-photon cascades from the
initial to the final states.
We found that those states with energies that lie between the 
energy of the initial and the final states, contribute a finite 
correction to the total two-photon decay rate. Taking into
account this correction, we calculate the rates for 
the $3S_{1/2} \to 1S_{1/2}$ two-photon decay of neutral hydrogen
as well as hydrogen-like ions. 
Our results are in a good agreement with nonrelativistic 
calculations for low $Z$ (see Refs.~\cite{Jen07,Jen07b})
but show a significant
deviation from the data by Cresser and co-workers 
\cite{CrT86}. 

Our quantum electrodynamics approach, as discussed in the 
present paper, opens a way for a \textit{systematic} theoretical
analysis of the simultaneous, coherent two-photon emission
from one-electron (and many-electron) atomic systems,
even in cases where problematic intermediate states with an energy 
between the initial and final states give rise to resonance
peaks in the photon energy distributions. 
We stress here that a conceivable alternative approach
to the removal of the formal infinities generated by the 
intermediate ``cascade'' states, which is based on the 
explicit removal of these states from the sums over $\nu$ and $\rho$
in Eq.~(\ref{gamma_2_QED}), gives rise to a number of conceptual 
problems, including gauge-noninvariance with respect to length and 
velocity gauges~\cite{Jen07b}. Our approach is manifestly gauge invariant
and also avoids problems connected with the identification of the infinitesimal
parts $i\epsilon$ in the propagator denominators in Eq.~(\ref{gamma_2_QED})
with partial or total decay rates of the intermediate states:
the $\epsilon$ parameters are free parameters which approach 
zero after all other operations, including the integrations
over the photon energies, have been performed. This operation 
leads to a finite result and corresponds, as explained in 
Ref.~\cite{Jen07b}, to a {\em partial} removal of the 
problematic intermediate states from the sum over all virtual 
states involved in the two-photon process, albeit in a fully
gauge-invariant manner. 

In addition to its relevance for atomic physics, our approach may have 
a significant impact for astrophysical studies where a detailed 
knowledge of the (properties of) two-photon transitions is highly 
required for the analysis of cosmological hydrogen and helium 
recombination. The contribution of two-photon processes to 
the recombination history represents an issue which 
has recently attracted substantial theoretical 
interest \cite{DuG05}.

\section*{ACKNOWLEDGMENTS}

U.D.J. acknowledges support from Deutsche Forschungsgemeinschaft (Heisenberg
program), and A.S. acknowledges support from the Helmholtz Gemeinschaft
(Nachwuchsgruppe VH--NG--421). The authors acknowledge insightful discussions
with Z. Harman regarding the Sturmian decomposition of the Dirac--Coulomb Green
function as given in Ref.~\cite{HyS97}, and helpful discussions with P.
Indelicato on general aspects of the two-photon decay and associated
resonances.

\end{document}